# Efficient Entity Resolution on Heterogeneous Records

Yiming Lin, Hongzhi Wang, Jianzhong Li, Hong Gao
Harbin Institute of Technology, Harbin, China
YimingLin_0426@hotmail.com {wangzh,lijzh,honggao}@hit.edu.cn

*Abstract*—Entity resolution (ER) is the problem of identifying and merging records that refer to the same real-world entity. In many scenarios, raw records are stored under heterogeneous environment. Specifically, the schemas of records may differ from each other. To leverage such records better, most existing work assume that schema matching and data exchange have been done to convert records under different schemas to those under a predefined schema. However, we observe that schema matching would lose information in some cases, which could be useful or even crucial to ER.

To leverage sufficient information from heterogeneous sources, in this paper, we address several challenges of ER on heterogeneous records and show that none of existing similarity metrics or their transformations could be applied to find similar records under heterogeneous settings. Motivated by this, we design the similarity function and propose a novel framework to iteratively find records which refer to the same entity. Regarding efficiency, we build an index to generate candidates and accelerate similarity computation. Evaluations on real-world datasets show the effectiveness and efficiency of our methods.

## I. INTRODUCTION

Entity Resolution (ER) is the process of identifying and merging records that refer to the same real-world entity across different data sources. It is a crucial step for data cleaning and data integration.

As surveyed in [1], due to the success of social and semantic web applications, as well as the establishment of standards and best practices for publishing and exchanging data on the Web, a large and quickly growing volume of heterogeneous datasets has become available on the Web. ER on heterogeneous data is in demand.

Most existing techniques studied ER on records under a predefined schema[2]. However, in many scenarios, the lineage of such records comes from heterogeneous sources, the schemas of which could vary from source to source. To convert records under different schemas to those under a predefined schema, two steps are often required, i.e. schema matching and data exchange. First, we perform schema matching, which is a specification that describes how a record structured under one schema (the source schema) is to be transformed into a record structured under a different schema (the target schema). Given the schema matching, we execute data exchange to convert an instance of the source schema into an instance of the target schema so that the schema matching is satisfied.

In the conventional framework shown in Fig 1-(c), schema matching and data exchange are performed to resolve semantic heterogeneity and transform records under a predefined schema, alleviating the need for extensive approaches of ER. Even though such framework works for many scenarios, it would *lose information* in some cases, which is useful or even crucial to entity resolution. Information loss arises from the difference of information content in target schema and source schemas. In most scenarios, a target schema is defined by the user for specific computation goals. In contrast, the source schema is created individually and often envelops a wide range of information describing entities. Thus, when the difference of information content between source schema and target schema is pronounced, information loss is an inevitable obstacle for entity resolution on records under target schema.

Both false positive and false negative errors are led by information loss. We use an example to illustrate such cases.

*Example 1:* Consider a company with three types of customer records shown in Fig 1, each of which owns its schema, respectively. Given a target schema {{name},{position}, {addr},{city}} and schema matchings, we execute data exchange to convert the instances form source schemas to the target schema shown in Fig 1-($b$). The ground truth can be obtained manually or by external knowledge such as knowledge bases or crowdsourcing. Note that the ground truth is not required for our proposed approach and only used to illustrate the example.

Now consider records under the target schema. Under various string metrics, such as edit distance, Jaccard similarity, cosine etc., the similarity of $r_7$ and $r_8$ is high. However, they do not refer to the same entity according to the ground truth. Under the target schema, it is infeasible to correct this false positive. Instead, if we explore information from records under source schemas, we find that $r_7$ is the join result of $r_2$ and $r_4$, and $r_8$ is the join result of $r_3$ and $r_5$. The information useful for entity resolution, e-mail, Tel/Contact No and Con.Type(Consumption type) are dropped during schema matching. From such distinguished information, we could infer that $r_7$ and $r_8$ refer to different entities.

Next consider false negative error. $r_7$ and $r_9$ share few similar fields under target schema but the ground truth indicates they refer to the same entity. To avoid false negative errors, we need to collect potential positive evidence from instances under source schema. That is, $r_9$ is the join result of $r_1$ and $r_6$. For $r_7$ and $r_9$, we examine e-mail and TeL/Contact No, which is identical (both are {bush@gmail, 831-432}). Also their Con.Type is high similar. Thus, it is possible to draw the conclusion that they describe the same entity.

This example shows that schema matching would lose information in the case when the information content between target schema and source schemas is different. Then could we make such transformations to avoid information loss? Such transformations have following two steps.

1) We first construct a schema (called *full schema*) containing all the attributes from different sources schemas and then convert their instances to those under full schema;

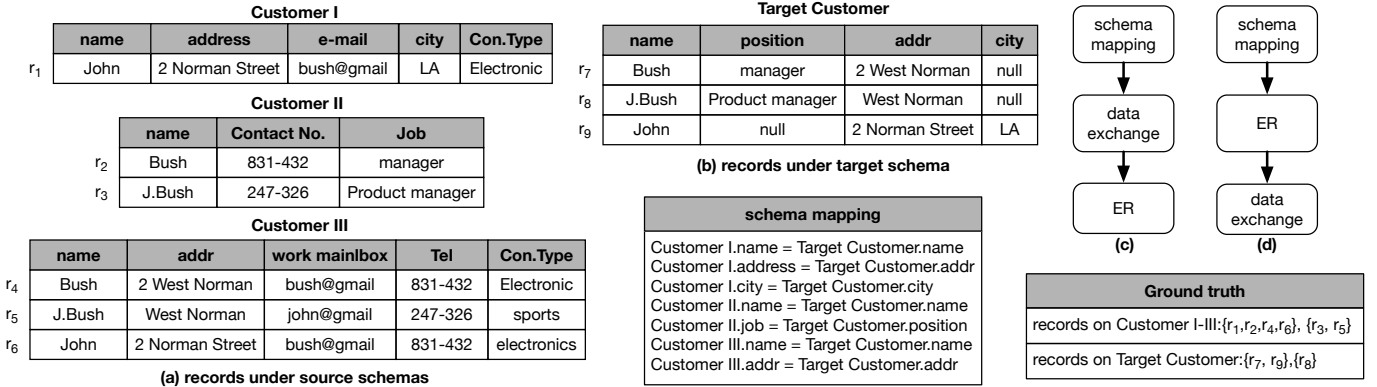

Fig. 1. Motivating Example: A Customer Mapping Scenario.

2) We perform ER on those records under full schema. Then we convert instances under full schema to those under target schema.

The above transformations provide a chance for ER technique on *homogeneous records* (records under the same schema) to be applied to solve the problem under heterogeneous settings. However, we claim the above transformations are infeasible in practice. The complete schema matchings between heterogeneous sources are difficult to capture since it is pretty costly to build pairwise matchings for schemas from massive data sources. Thus, records under full schema will contain considerable *null* attribute values, which could lower the quality of those records significantly.

In summary, ER would lose information on homogeneous records in the case when the information content between source schemas and target schema is different. To our best knowledge, none of existing technique of ER on homogeneous records or its transformation could solve this problem. When the lost information is discriminative, it would probably lower the quality of results for ER on records under target schema. We highlight that the impaired quality resulted from information loss **CANNOT** be offset by the advance in ER techniques on homogeneous records since we can hardly tackle a record losing information, which may be essential, correctly.

Motivated by Example 1 and the above discussions, we seek to leverage sufficient information from heterogeneous sources to improve ER instead of only the attributes in the target schema. Thus, we propose a new framework as shown in Fig 1-(d). Given the schema matchings between source schemas and target schema, we perform ER on heterogeneous records directly. Finally, we execute data exchange to convert heterogeneous records with entity labels to a predefined schema (target schema). Note that both frameworks shown in Fig 1-(c) and (d) finally produce records with homogeneous schema (target schema), each of which is with an entity label.

Such a framework is advantageous in two aspects.

On one hand, for entity resolution on homogeneous records after schema matching and data exchange, information loss is irreparable and sometimes unavoidable, which will reduce the quality of ER. Sailing information in heterogeneous sources enables us to find all potential positive evidences to improve ER quality.

On the other hand, ER will improve the quality of data exchange. An ideal data exchange is to join instances referring to the same real-world entity. However, most existing work about data exchange join two records with the same or similar key values, which may not represent that the two instances refer to the same real-world entity. To solve this problem, our framework accomplishes ER before data exchange, which offers feasibility to an ideal exchange. Since in this paper we focus on ER, the chances of improvement for data exchange are discussed in [3].

Even though ER directly on records in heterogeneous schemas brings some advantages, new challenges are posed.

First, two records referring to the same real-world entity may describe its different aspects with different attributes. As an example, consider $r_1$ and $r_2$ in Fig 1. Ground truth indicates that they refer to the same entity, but they only share one attribute name and the other 6 attributes are different. Thus, their similarity could be low. We capture this phenomenon as *description difference*, which make it non-trivial to match such heterogeneous record pairs correctly.

Second, when comparing two records, the insufficiency of schema matching between source schemas leads to the difficulty in similarity computation. As an instance, for $r_1$ and $r_4$, since we are not sure whether Customer I.email and Customer III.work mailbox refer to the same attribute of one entity, we could not compare corresponding attribute values directly. We capture this challenge as *heterogeneous schema*. None of similarity function designed for records under identical schema [2] can be used to compute the similarity of two heterogeneous records directly.

To our best knowledge, the critical feature *description difference* is first defined by us. Furthermore, none of existing work on ER could handle above two challenges at the same time. Most work assumes that schema matching and data exchange have been accomplished to make the records from different sources be compared in a uniform manner [4], [5], [6]. They focus on whether records refer to the same entity under a given specific schema, which would possibly neglect some crucial positive evidence for ER such as some distinguished attributes excluded by target schema. Regarding human-based approaches (crowd source) [7], [8], we believe that for those records satisfying *description difference*, human may probably return wrong answers: such as $r_1$ and $r_2$ in the motivating example, they are easily mistaken for belonging to different entities. (But actually they link to the same one) There are also some learning-based approaches studied ER on heterogeneous records. [9] applied classification techniques drawn from statistical pattern recognition, machine learning to decide whether two records represent the same real-world entity. [10] presented a algorithm *HFM* that combines the machine learning and expert systems approaches to determine true field matchings. [9], [10] both employed a supervised learning which requires training datasets with ground truth. However, the extraction of such datasets is very difficult and expensive since those record pairs with *description difference* are hard to be recognized even by human, not to mention generating a sufficient training data. Comparing with them, we can perform ER without training datasets and any aprior information.

To solve the above two challenges, we propose HERA (*Heterogeneous Entity Resolution Algorithm*). HERA could handle records with various data types, such as string data, numeric data, etc. and view the similarity metric of corresponding data type as a black-box, which permits extensive ER resolution. Regarding *description difference*, HERA uses a compare-and-merge mechanism to iteratively find and merge similar records. For *heterogeneous schema*, HERA could compute the similarity of two records without any priori schema matchings. Furthermore, HERA can generate some high-reliable schema matchings to help record similarity computation. We also take the efficiency issue into account and design an efficient index. Based on it, we derive a tight upper bound and lower bound of record similarity to generate candidates within linear time. Leveraging index, the time complexity of record similarity computation is reduced by three orders of magnitude comparing with a basic nest-loop method and two records merging can be accomplished within logarithmic steps.

In summary, we make the following contributions:

- We propose a novel and extensive framework for ER under heterogeneous environment. To support such framework, we define a *super record* as the merged record of those referring to one entity. It provides more evidence for entity resolution on records with heterogenous schemas.
- Based on the framework, we propose HERA to handle ER on heterogenous records. In this algorithm, we develop an efficient index, an iterative compare-and-merge mechanism, bipartite-based heterogenous record matching approach and schema matching prediction method based on probabilistic majority voting. These novel techniques ensure the efficiency and effectiveness of HERA.
- We conduct a comprehensive evaluation of our techniques against real-world data sets. Our experimental results show that HERA achieves high performance and could collect valid positive evidence to improve the results of ER with a predefined schema significantly.

In the rest of the paper, we formally formulate ER on heterogeneous records and give an overview in Section II. Next we design the index in Section III and propose approaches for record similarity computation in Section IV. Section V presents the overall solution for ER. Finally, in Section VI we reports the experimental results with analysis; Section VII discusses the related work; and Section VIII concludes the whole paper.

## II. PROBLEM STATEMENT AND OVERVIEW

In this section, we define the problem and related concepts in Section II-A. Then, in Section II-B we overview our solution.

### A. Definitions

For a record set $\mathcal{R} = \{r_1, r_2, ..., r_n\}$ with heterogenous schemas, the schema of $r_i$ is $s_i$ with $k_i$ attributes, $a_1^i, a_2^i, \cdots,$ $a_{k_i}^i$. We denote by $a_t^i \approx a_l^j$ if $a_t^i$ is mapped to $a_l^j$. The problem of entity resolution is defined as follows.

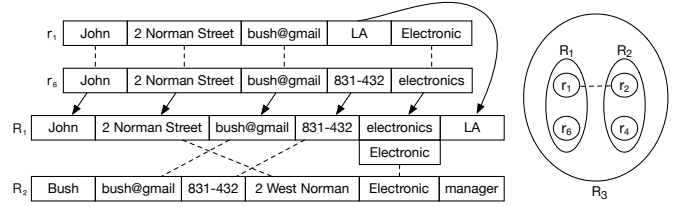

Fig. 2. Super record

*Definition 1:* (PROBLEM) Given a record set $\mathcal{R}$ with heterogenous schemas , identify and merge records that refer to the same real-world entity. (ER)

As discussed in Section I, ER on heterogenous data brings the new challenge of description difference. We say two records are in *description difference*, if they refer to the same entity but share few attributes, and thus have a low similarity. Traditional approaches based on similarity functions can hardly handle such case. Facing this challenge, we propose *Compare-and-Merge* mechanism. That is, we merge the records that are determined as referring to the same entity into a *super record*, and perform entity resolution on super records instead of computing similarities between records directly. Super records extend the information of the entity that it refers to. They provide more evidence to identify that a record refers to an entity. Continue with motivating example in Fig 1. The ground truth indicates that $r_1$ and $r_2$ refer to the same entity. However, we cannot merge them directly since their similarity is low. As shown in Fig 2, we observe that $r_1$ and $r_6$ are similar, thus we merge them into a super record $R_1$; $r_2$ and $r_4$ are similar, then we merge them into a super record $R_2$. Also, the similarity between $R_1$ and $R_2$ is high, thus we merge $R_1$ and $R_2$ into $R_3$. Now, $r_1$ and $r_2$ have been merged into one super record. As a result, super records could be used to resolve description difference.

The structure of a super record is defined as follows.

*Definition 2:* (SUPER RECORD) A super record $R = \{f_1^R, f_2^R, ..., f_{|R|}^R\}$, where $f_i^R$ is the set of values corresponding to the $i^{th}$ field of $R$. $f_i^R = \{v_{1,i}^R, v_{2,i}^R, ..., v_{|f_i^R|,i}^R\}$, where $v_{j,i}^R$ is the $j^{th}$ value of $f_i^R$.

In the remainder of this paper, if the location of a value or field in the record is not sensitive, then we abuse the notion. That is, let $v^R$ and $f^R$ be a value $v$ and a field $f$ of $R$.

Note that super record is a general representation of records. A basic record is the simplest super record, where each field stores one value. Without loss of generality, if $R_1$ and $R_2$ are merged into $R_3$, we denote that by, $R_3 = R_1 \oplus R_2$. Next we illustrate such merging operation as follows.

*Example 2:* Considering the motivating example, we seek to merge $r_1$ and $r_6$. As shown in Fig 2, first we merge corresponding fields of them, and then store multiple values for the same field (if any). For attribute Con.Type, $r_1$ and $r_6$ own different values {electronics}, {Electronic}, and we store both of them. For the other fields without matching relationship, such as LA, 831-432, we add them into super record directly.

Next, we devise the similarity of two super records, which is non-trivial due to its sophisticated structure. A super record consists of several fields, thus we first deduce the similarity of field pairs, and then define the record similarity.

According to Definition 2, a field stores multiple values which refer to the same attribute of an entity. If two fields are similar, they must share some similar values. Thus we use the similarity of the most similar value pair to evaluate field similarity, which is defined as follows.

*Definition 3:* (FIELD SIMILARITY) Given two super records $R_i, R_j$, let $simf(f_k^{R_i}, f_l^{R_j})$ be the similarity of two fields $f_k^{R_i}, f_l^{R_j}$, $simv(v_{p,k}^{R_i}, v_{q,l}^{R_j})$ be the similarity of two values $v_{p,k}^{R_i}, v_{q,l}^{R_j}$. That is,

$$simf(f_k^{R_i}, f_l^{R_j}) = \max_{1 \leq p \leq |f_k^{R_i}|, 1 \leq q \leq |f_l^{R_j}|} \{simv(v_{p,k}^{R_i}, v_{q,l}^{R_j})\}$$

Our approach could handle various data types, such as string data, numeric data, etc. And we view the corresponding similarity functions as black-box, which permits extensive ER solutions. Both typographical and semantics variations of data could be handled by our approach by involving special similarity function. For ease of illustration, in the motivating examples throughout our paper, we consider string data, which is most widely used, and take *Jaccard* as similarity metric, which is defined as follows.

Regarding two strings $v_1$ and $v_2$, let $\mathfrak{V}_1$ and $\mathfrak{V}_2$ be their corresponding q-grams set, respectively. Their Jaccard similarity $simv(v_1, v_2) = \frac{|\mathfrak{V}_1 \cap \mathfrak{V}_2|}{|\mathfrak{V}_1 \cup \mathfrak{V}_2|}$.

Note that other string similarity functions, such as *Soft TF-IDF*, edit distance, etc, could be served as alternatives.

Since we perform ER on records with heterogeneous schema directly, when deciding the similarity of two records, the matchings of their corresponding schemas is unknown. In Section IV we present the solution for finding such schema matchings. Here we assume that the matching information is available and give the following definition.

*Definition 4:* (FIELD MATCHING) Given two super records $R_i, R_j$ and a value similarity threshold $\xi$, if $f^{R_i}$ and $f^{R_j}$ are decided to be identical field of an entity, then $f^{R_i} \simeq f^{R_j}$, denoting that $f^{R_i}$ and $f^{R_j}$ are matched. Let $\mathcal{F}(i,j)$ be the field matching set. $\mathcal{F}(i,j) = \{(f^{R_i}, f^{R_j}) \mid f^{R_i} \simeq f^{R_j}, simf(f^{R_i}, f^{R_j}) \geq \xi\}$.

Intuitively, the similarity between two records $R_i$ and $R_j$ is the accumulation of the similarity between corresponding fields. In the case that $|R_i| < |R_j|$, when most of attributes in $R_i$ are similar to some attributes in $R_j$, $R_i$ could be considered similar as $R_j$, even though it can hardly find similar attributes in $R_i$ for some of attributes in $R_j$. With this consideration, to normalize the similarity between two records within [0,1] interval, we divide the accumulation by $min(|R_i|, |R_j|)$. Thus, the similarity between two records is defined as follows.

*Definition 5:* (RECORD SIMILARITY) Given two super records $R_i, R_j$, their similarity is defined as follows,

$$Sim(R_i, R_j) = \frac{\sum_{(f^{R_i}, f^{R_j}) \in \mathcal{F}(i,j)} simf(f^{R_i}, f^{R_j})}{min(|R_i|, |R_j|)} \quad (1)$$

We use an example to illustrate the similarity metrics above.

*Example 3:* Continuing with the motivating example, we consider two super records $R_1 = r_1 \oplus r_6$, $R_2 = r_2 \oplus r_4$, as shown in Fig 2. First, regarding field similarity, for attribute Con.Type, $f_5^{R_1} = \{\{Electronic\}\}, \{electronics\}\}$, $f_5^{R_2} = \{\{Electronic\}\}$. Since $simv(\{Electronic\}, \{Electronic\}) > simv(\{electronics\}, \{Electronic\})$ (we set 2 q-grams), $simf(f_5^{R_1}, f_5^{R_2}) = simv(\{Electronic\}, \{Electronic\}) = 1$.

The dotted lines represent field matchings. If we set $\xi = 0.35$, then the field matching set $\mathcal{F}(1,2) = \{(f_2^{R_1}, f_4^{R_2}), (f_3^{R_1}, f_2^{R_2}), (f_4^{R_1}, f_3^{R_2}), (f_5^{R_1}, f_5^{R_2})\}$. By accumulating field similarity of field pair in $\mathcal{F}(1,2)$ and dividing it by 6, we obtain the record similarity of $R_1$ and $R_2$, $Sim(R_1, R_2) = \frac{0.37 + 1.0 + 1.0 + 1.0}{6} = 0.56$.

*B. Overview*

In this part, we overview our solution for ER on heterogeneous records.

Due to *description difference* arisen from heterogeneous settings, those records could hardly be detected by batch processing, and we propose *merge-and-compare* mechanism to handle description difference. That is, we merge any similar records into a super record. When deciding whether two records $r_i$ and $r_j$ are similar, instead of computing their record similarity straightforward, we consider the similarity of their corresponding super records. If their super records are similar, we deem that $r_i$ and $r_j$ refer to the same entity. Then we merge their corresponding super records to find potential new matchings in later rounds.

To facilitate above process, we adopts an iterative method to find and merge similar records. The stop condition is that the similarity of any two super records is below a predefined threshold. In each iteration, we first generate candidate record pairs and then verify them. These two steps are introduced as follows.

**Candidate Generation** This step prunes dissimilar record pairs and ensures high similar record pairs as the candidates.

According to Definition 5, if two records are similar, they must share some similar values. With this consideration, in this step, we aim to obtain the record pairs with at least one shared similar value. To achieve this goal, we index similar value pairs, such that a set of record pairs sharing at least one similar value could be obtained as the candidate set within linear time. Then, to refine the candidate set, for each candidate record pair, we derive an upper bound and a lower bound of the record similarity with logarithmic steps. That is, given a record similarity threshold $\delta$, for a record pair $(R_i, R_j)$, if the upper bound of $Sim(R_i, R_j)$ is below $\delta$, then we could prune this pair safely; if the upper bound of $Sim(R_i, R_j)$ is equal to its lower bound, then we could directly determine the record similarity without verification.

**Verification** For each candidate record pair, we aim to verify it by judging whether its similarity exceeds the threshold. The core of this step is to compute the similarity of two heterogenous super records.

Such similarity computation is non-trivial due to two-fold aspects: schema matching information between two records is missing and the efficiency need to be taken into account owing to the complex structure of super records.

In absence of the schema matchings of records, we seek to derive similarity by detecting the similar degree of their values. Thus, we ignore the attributes information and treat a record as a set of values to derive record similarity, which is captured by *instance-based* method.

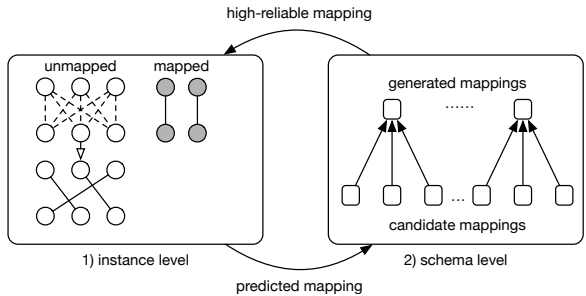

Fig. 3. Interaction of instance-based and schema-based method

For two similar records, if two fields are high similar, their corresponding attributes could potentially match. Thus, a similar record pair would provide some predictions for corresponding schema matchings. We could use such predictions to generate some high-reliable schema matchings, which is captured by *schema-based* method.

With this consideration, we combine both *instance-based* and *schema-based* method for similarity computation because they can benefit each other. As shown in Fig 3, instance-based method can provide predictions of schema matching, which leads schema-based method to determine more accurate matchings. With the high-reliable matchings generated by schema-based method, we could compute the similarity of two mapped fields straightforward, which can accelerate record similarity computation and improve the accuracy.

Next, we introduce these two methods briefly.

- *instance-based method:* The core part of record similarity computation is to determine the field matching set.
  According to Definition 4, basically, we first need to find all similar field pairs. This operation can be transformed to performing similarity join on two value sets of the record pair, which could be facilitated efficiently by index. Then, we model the problem of deciding field matching set as finding a maximum weight matching in bipartite graph, which is solved by Kuhn CMunkres [11] and sped up by graph simplification algorithms.
- *schema-based method:* Among predictions provided by instance-based method, one attribute may match multiple attributes. Consider a common assumption[12] that there are no redundant attributes in one schema. Thus, the determination of the matching relationship for such one-to-many matching is converted to the problem of finding the proper matching attribute from multiple choices. A probabilistic method based on majority vote could estimate the matching attribute with the upper bound of error probability. The details will be introduced in Section IV-B.

In summary, in the remainder of our paper, Section III presents the details of index building and candidate generation. Verification step would be described in Section IV.

## III. INDEX

In Section II-B, we overview the solution and show that index is required to achieve two goals. One is to generate candidate record pairs. The other is to accelerate record similarity computation. To achieve these goals, in this section, we introduce the index, whose optimization relies on the fact that the record similarity is essentially a combination of some value similarities. Thus, our goal is first to find similar value pairs between records efficiently leveraging index (Section III-A). And then we describe how to reach the proposed optimization objectives. (Section III-B).

### A. Index Structure

Clearly, if two records share a lot of similar value pairs, they could be potentially similar. To generate proper candidate record pairs, we require to give a fast estimation of their similarity score using index efficiently. Thus, we index value pairs with similarity above the given value similarity threshold.

To begin with, for each value $v_{k,j}^{R_i}$ in $\mathcal{R}$, we assign it a unique label $(rid, fid, vid)$, where $rid = i$, $fid = j$, $vid = k$. And we assume that all the labels in our index begin from 1. In Fig 2, the label of {Electronic} in $R_1$ is $(1, 5, 2)$ and that of {bush@gmail} in $R_2$ is $(2, 2, 1)$.

*Definition 6:* (INDEX) Let $\mathcal{V}$ be the set of value pairs stored in index. For each value pair $(v_{k,p}^{R_i}, v_{l,q}^{R_j}) \in \mathcal{V}$ where $i \neq j$, we index it by assigning each value pair a label: $((i,p,k),(j,q,l))$. Next we execute the following operations:

1) For a value pair with label $((rid_1, fid_1, vid_1), (rid_2, fid_2, vid_2))$, exchange their locations to ensure $rid_1 < rid_2$;

2) Value pairs in $\mathcal{V}$ are sorted in the priority of $rid_1$, $rid_2$ and the value similarity. That is, value pairs are sorted according to $rid_1$ and $rid_2$ in ascending order. Those pairs with same $rid_1$ and $rid_2$ are sorted by the value similarity in descending order.

Each value pair in index owns a $pid$, denoting $pid^{th}$ pair.

In summary, for each value pair in index, we stored its $pid$, corresponding labels and value similarity.

As an example, the index for value pairs in motivating example ($r_1$ to $r_6$) is shown in Fig 4. The part contained in the solid rectangle represents the index and contains totally 17 value pairs. Taking $13^{th}$ value pair for instance, we store its $pid$ as 13, label as $((4,1,1),(5,2,1))$ and its similarity score as 0.83. From its label, we can find its corresponding record pair as $(r_4, r_5)$. Noting that $rid_1$ and $rid_2$ (covered in dotted rectangle), $rid_1$ is sorted in ascending order. If two pairs share the same $rid_1$, they are sorted in ascending order by $rid_2$. Also, $13^{th}$ and $14^{th}$ pair share the same $rid_1$ and $rid_2$. Thus, they are sorted in the descending order by similarity score.

To build the index, we require to find all similar value pairs. Specifically, they satisfy both following two conditions:

1) their value similarity is above $\xi$;

2) they belongs to different records in $\mathcal{R}$;

This problem could be solved by *similarity join* [13] defined as follows.

*Definition 7:* (SIMILARITY JOIN) Given a record set $\mathcal{R}$, let $\mathbb{V}$ be value set present in $\mathcal{R}$, the result of *similarity join* on $\mathbb{V}$ is a set of value pairs $\mathcal{V}$,

$$\mathcal{V} = \{(v^{R_i}, v^{R_j}) \mid 1 \leq i,j \leq |\mathcal{R}|, i \neq j, simv(v^{R_i}, v^{R_j}) \geq \xi\}$$

Note that similarity join is not limited to string data type. In [3], we discuss the transformations of current *similarity join* techniques to apply to various data types.

Lei $\mathbb{V}_i$ and $\mathbb{V}_j$ be the set of values in records $R_i$ and $R_j$, respectively. We denote by $\mathcal{V}_{ij}$ the results of similarity join on

| pid | label | similarity | record pair |
|---|---|---|---|
| 1 | ((1,3,1), (4,3,1)) | 1 | (r1,r4) |
| 2 | ((1,1,1), (6,1,1)) | 1 | (r1,r6) |
| 3 | ((1,2,1), (6,2,1)) | 1 | (r1,r6) |
| 4 | ((1,3,1), (6,3,1)) | 1 | (r1,r6) |
| ... | ...... | ...... | ...... |
| 13 | ((4,1,1), (5,2,1)) | 0.83 | (r4,r5) |
| 14 | ((4,2,1), (5,1,1)) | 0.6 | (r4,r5) |
| 15 | ((4,3,1), (6,3,1)) | 1 | (r4,r6) |
| 16 | ((4,4,1), (6,4,1)) | 1 | (r4,r6) |
| 17 | ((4,5,1), (6,5,1)) | 0.9 | (r4,r6) |

(a) pid  (b) label  (c) similarity  (d) record pair

Fig. 4. Structure of Index

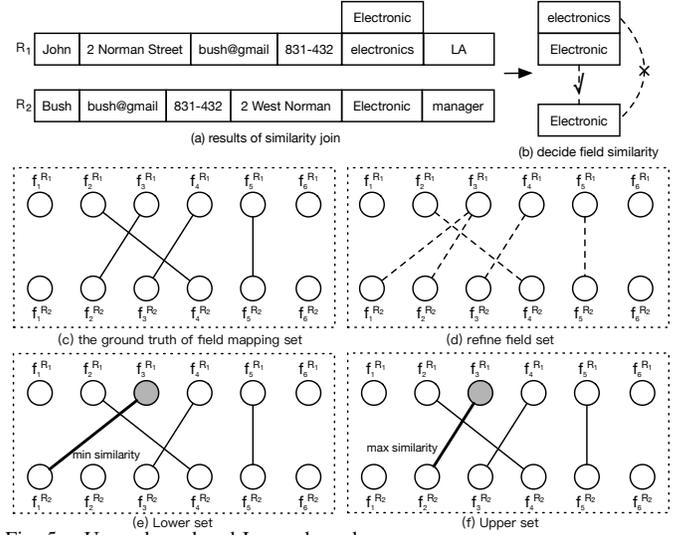

Fig. 5. Upper bound and Lower bound

$\mathbb{V}_i$ and $\mathbb{V}_j$.
$$\mathcal{V}_{ij} = \{(v^{R_i}, v^{R_j}) \mid simv(v^{R_i}, v^{R_j}) \geq \xi\}$$

Note that $\forall i,j \in [1, |\mathcal{R}|]$, we have $\mathcal{V}_{ij} \subseteq \mathcal{V}$, i.e. we can obtain similar value pairs of any two records from index. Thus, similarity join on $\mathbb{V}$ requires only one pass. That is to say, the index could be built off-line.

*Proposition 1:* Time complexity of index construction is $O(|\mathcal{V}|log|\mathcal{V}|)$.

Proof is shown in [3].

### B. Index Operation

In this section, we describe the operations on index to reach the aforementioned optimization goals and then present the details of index update arising from record merging.

*1) Candidate Generation:* As discussed in Section V, the goal of this operation is to filter dissimilar record pairs and determine the similarity of some records directly. These two parts are excluded from the candidate set. Let $\mathbb{R}$ be the candidate set and $\mathbb{R}'$ be the set of records whose similarity could be directly computed by index, respectively. Dissimilar pairs will not be considered furthermore, while the pairs in $\mathbb{R}'$ are included in the final results without further computation.

The results of this step are *candidate* record pairs, which is for further verification. Accordingly, given a record similarity threshold $\delta$, for $(R_i, R_j)$, we deduce an upper bound and a lower bound of record similarity, denoted by $Up(R_i, R_j)$ and $Low(R_i, R_j)$.

If $Up(R_i, R_j) < \delta$, then $(R_i, R_j)$ could be safely pruned since $Sim(R_i, R_j) < \delta$ must hold. If $Up(R_i, R_j) = Low(R_i, R_j)$, then we could directly determine $Sim(R_i, R_j) = Up(R_i, R_j)$.

According to Definition 5, the key of computing $Sim(R_i, R_j)$ is to decide a field matching set $\mathcal{F}_{ij}$, which consists of several matching field pairs following one-to-one matching. As an example in Fig 5, we seek to deduce bounds of $Sim(R_1, R_2)$ and $\xi$ is set as 0.3. We view each field as a point and Fig 5-(c) depicts the ground truth of field matching $\mathcal{F}_{12}$. Next we use $\mathcal{V}_{ij}$ to derive $Up(R_i, R_j)$ and $Low(R_i, R_j)$ as follows.

1) For each field pair, we store the value pair with the maximum similarity and delete others. We denote such field pair set after the above deletions by $\mathcal{V}'_{ij}$, which is called the *refined field set*. As shown in Fig 5-(b), for field pair $(f_5^{R_1}, f_5^{R_2})$, we retain {{Electronic},{Electronic}} and delete others since its similarity is maximum. Fig 5-(d) shows $\mathcal{V}'_{12}$.

2) Regarding $\mathcal{V}'_{ij}$, if a field is only contained in one field pair, we call it a *single field*; if a field is covered by more than one field pairs, we call it a *multiple field*. $f_3^{R_1}$ in Fig 5-(d) is a *multiple field* since it is covered by $(f_3^{R_1}, f_1^{R_2})$ and $(f_3^{R_1}, f_2^{R_2})$. Next consider two cases: For each *multiple field*, among the field pairs covering it, we retain the one with the maximum similarity and delete others. $\mathcal{V}'_{ij}$ after such operations is called an *Upper set*, which is denoted by $\mathcal{US}'_{ij}$ (shown in Fig 5-(f)); For each *multiple field*, among the field pairs covering it, we retain the one with minimum similarity and delete others. $\mathcal{V}'_{ij}$ after such operations is called a *Lower set*, which is denoted by $\mathcal{LS}'_{ij}$ (shown in Fig 5-(e)). Note that for each *multiple field*, among the field pairs covering it, only one pair is contained in $\mathcal{F}_{ij}$ since the pairs in $\mathcal{F}_{ij}$ follow one-to-one matching. And how to decide such a pair will be described in Section IV-A in detail.

Thus, the set of field pairs that do not cover any *multiple field* is the intersection of $\mathcal{F}_{ij}$, $\mathcal{US}'_{ij}$ and $\mathcal{LS}'_{ij}$. Except them, for each *multiple field*, $\mathcal{US}'_{ij}$ covers the pair with the maximum similarity and $\mathcal{LS}'_{ij}$ covers the pair with the minimum similarity. Therefore, the sum of field similarities for the above three sets satisfies the following inequalities.

$$\sum_{(f^{R_i}, f^{R_j}) \in \mathcal{LS}'_{ij}} simf(f^{R_i}, f^{R_j}) \leq \sum_{(f^{R_i}, f^{R_j}) \in \mathcal{F}(i,j)} simf(f^{R_i}, f^{R_j})$$
$$\leq \sum_{(f^{R_i}, f^{R_j}) \in \mathcal{US}'_{ij}} simf(f^{R_i}, f^{R_j}) \quad (2)$$

According to Definition 5, the upper bound and lower bound of $Sim(R_i, R_j)$ is,

$$Up(R_i, R_j) = \frac{\sum_{(f^{R_i}, f^{R_j}) \in \mathcal{US}'_{ij}} simf(f^{R_i}, f^{R_j})}{\min(R_i, R_j)} \quad (3)$$

$$Low(R_i, R_j) = \frac{\sum_{(f^{R_i}, f^{R_j}) \in \mathcal{LS}'_{ij}} simf(f^{R_i}, f^{R_j})}{\min(R_i, R_j)} \quad (4)$$

Typically, if *multiple field* does not exist in $R_i$ or $R_j$, we have $\mathcal{F}_{ij} = \mathcal{US}'_{ij}$ and $\mathcal{F}_{ij} = \mathcal{LS}'_{ij}$. In this case, the lower bound is equal to upper bound, thus record similarity is exactly the upper or lower bound. For such record pairs, we exclude them from candidate set and merge them directly. The details of merge operation is presented in Section III-B2.

As an example in Fig 5, $\mathcal{US}'_{12} = \{(f_2^{R_1}, f_4^{R_2}), (f_3^{R_1}, f_2^{R_2}), (f_4^{R_1}, f_3^{R_2}), (f_5^{R_1}, f_5^{R_2})\}$, $\mathcal{LS}'_{12} = \{(f_2^{R_1}, f_4^{R_2}), (f_3^{R_1}, f_1^{R_2}), (f_4^{R_1}, f_3^{R_2}), (f_5^{R_1}, f_5^{R_2}), \}$. $f_3^{R_1}$ is the only *multiple field*. $\mathcal{US}'_{12}$ contains the field pair $(f_3^{R_1}, f_2^{R_2})$ with the maximum similarity, i.e. (bush@gmail,bush@gmail). While $\mathcal{LS}'_{12}$ contains (bush@gmail,Bush) with minimum similarity. Thus, $Up(R_i, R_j) = \frac{0.37+1+1+1}{6} = 0.56$, $Low(R_i, R_j) = \frac{0.37+0.33+1+1}{6} = 0.45$.

Next we illustrate how to compute $Up(R_i, R_j)$ and $Low(R_i, R_j)$ through index in Algorithm 1. For convenience of expression, each value pair is expressed as $((rid_1, fid_1, vid_1), (rid_2, fid_2, vid_2))$ and the $pid^{th}$ value pair is denoted by $pair^{pid}$, whose similarity is $simv_{pid}$.

---

**Algorithm 1** Cal_bound(i,j)

**Input:** $\mathcal{V}, i, j$;
**Output:** $Up(R_i, R_j), Low(R_i, R_j)$;
1: $sim \leftarrow 0, id \leftarrow 0$; // $Sim$ to record similarity, $id$ to record corresponding $pid$;
2: $flagU \leftarrow false, flagL \leftarrow false, pre \leftarrow 0$;
3: // To find $\mathcal{V}'_{ij}$;
4: $(k, l) \leftarrow binary\_search\_l(1, |\mathcal{V}|, i)$;
5: $(p, q) \leftarrow binary\_search\_r(k, l, j)$;
6: **for** $p \leq pid \leq q$ **do**
7:    **if** $sim_{pid} > sim[fid_1][fid_2]$ **then**
8:      $\mathcal{V}'_{ij} \leftarrow \mathcal{V}'_{ij} \cup pair^{pid}, \mathcal{V}'_{ij} \leftarrow \mathcal{V}'_{ij} \setminus pair^{id[fid_1][fid_2]}$, $sim[fid_1][fid_2] \leftarrow sim_{pid}, id[fid_1][fid_2] \leftarrow pid$;
9: //So far we have computed $\mathcal{V}'_{ij}$. Next find $\mathcal{US}'_{ij}$ and $\mathcal{LS}'_{ij}$;
10: **for** each $pair^{pid} \in \mathcal{V}'_{ij}$ **do**
11:    **if** $!flagU[rid_1][fid_1]$ **then**
12:      $flagU[rid_1][fid_1] \leftarrow true, \mathcal{US}'_{ij} \leftarrow pair^{pid}$;
13:    **if** $!flagL[rid_1][fid_1]$ **then**
14:      $flagL[rid_1][fid_1] \leftarrow true$;
15:      **if** $pre \neq 0$ **then**
16:        $\mathcal{LS}'_{ij} \leftarrow pair^{pre}$;
17:    $pre \leftarrow pid$;
18: Compute $Up(R_i, R_j), Low(R_i, R_j)$ according to Equation 3 and 4.
19: **return** $Up(R_i, R_j), Low(R_i, R_j)$;

---

In Algorithm 1, we set two arrays $sim$ and $id$. $sim[i][j]$ is to record the maximum similarity of a field pair $(f^{R_i}, f^{R_j})$ so far and $id$ is to record the pid of corresponding value pair (Line 1). $flagU$, $flagL$ and $pre$ is used to find $\mathcal{US}'_{ij}$ and $\mathcal{LS}'_{ij}$ (Line 2). First, we seek to find $\mathcal{V}'_{ij}$ (Line 4-8). $binary\_search\_l(1, |\mathcal{V}|, i)$ searches the first and the last occurrences of $i$ as the value of $rid_1$ between the first and the $|\mathcal{V}|^{th}$ entries in $\mathcal{V}$. This function is implemented by binary search since $\mathcal{V}$ is sorted by $rid_1$. $binary\_search\_r(k, l, j)$ searches the first and the last occurrences of $j$ as the value of $rid_2$ in between $k^{th}$ and $l^{th}$ entries in $\mathcal{V}$. This function is also implemented by binary search since $\mathcal{V}$ is sorted by $rid_2$ when $rid_1$ of value pair is same. After that, the set of value pairs whose $pid \in [p, q]$ corresponds to $\mathcal{V}_{ij}$. Next, for each field pair, we only store the value pair with maximum similarity and delete others to obtain $\mathcal{V}'_{ij}$ (Line 7-8).

Then, we find $\mathcal{US}'_{ij}$ and $\mathcal{LS}'_{ij}$ (Line 10-17). For each field in $R_i$, we select the field pair covering it with maximum similarity. Since value pairs are sorted in descending order of its similarity value when rids are same, value pair we first meet is the one with maximum similarity (Line 11-12). For each field in $R_i$, we select the pair covering it with minimum similarity (Line 13-17). Finally, $Up(R_i, R_j)$ and $Low(R_i, R_j)$ are computed (Line 18).

*Example 4:* In Fig 4, the index is enveloped in a solid rectangle. If we attempt to find $V'_{46}$, we first locate 4 using binary search as $rid_1$ for each value pair, which are enveloped in a dotted rectangle. We find 4 (in bold) appears from $13^{th}$ to $17^{th}$ value pair. Then we binary search $13^{th}$ to $17^{th}$ value pairs to find the entry with $rid_2 = 6$. Finally, we find three value pairs, $\{((4, 3, 1), (6, 3, 1)), ((4, 4, 1), (6, 4, 1)), ((4, 5, 1), (6, 5, 1))\}$.

Next we deduce $Up(r_4, r_6)$ and set the record similarity threshold $\delta$ as 0.5. $Up(r_4, r_6) = \frac{1+1+0.9}{\min(5,5)} = 0.58$, $Low(r_4, r_6) = \frac{1+1+0.9}{\min(5,5)} = 0.58$, both of which are same because there are no *multiple field* for $(r_4, r_6)$. Thus, $(r_4, r_6)$ is not a candidate pair since its similarity is computed directly by index.

In total, $\binom{2}{6} = 15$ record pairs are generated from 6 records. Using the upper and lower bound constraint, there is only one candidate, $\{(r_2, r_4)\}$.

*Proposition 2: Candidate generation* requires $O(|\mathcal{V}|)$ steps.

We show the proof in [3].

*2) Index Maintenance:* To resolve *description difference* discussed in Section I, once two records are judged to be similar, we merge them into a super record. Since merging two records would change the labels of corresponding value pairs, such operation may involve further modification of index structure and is not trivial. Thus, in this section, we discuss how to maintain index when merging two records efficiently.

Without loss of generality, we consider merging $R_i$ and $R_j$ into $R_k$. The process is as follows.

1) *merge:* We use the union-find structure [14] to maintain $rid$ (The functions of unoin and find are shown in [3]), i.e.,

---

((1,3,1), (4,3,1)) ⟶ ((1,3,1), (4,3,1))

((1,1,1), (6,1,1)) ⟶ ((1,1,1), (6,1,1))

((1,2,1), (6,2,1)) ⟶ ((1,2,1), (6,2,1))

((1,3,1), (6,3,1)) ⟶ ((1,3,1), (6,3,1))

((1,5,1), (6,5,1)) ⟶ ((1,5,1), (6,5,1))

((2,2,1), (6,4,1)) ⟶ ((2,2,1), (1,4,1))

((4,3,1), (6,3,1)) ⟶ ((4,3,1), (1,3,1))

((4,4,1), (6,4,1)) ⟶ ((4,4,1), (1,4,1))

((4,5,1), (6,5,1)) ⟶ ((2,2,1), (1,5,1))

Fig. 6. Label update by merge operation

k=unoin(i,j). Next, for each field pair $(f^{R_i}, f^{R_j}) \in \mathcal{F}_{ij}$ (see Definition 4), we merge $f^{R_i}$ and $f^{R_j}$ into one field $f^{R_k}$ and then merge their corresponding values. Thus, each value in $R_k$ would be assigned a new label.

2) *delete:* For those value pairs between $R_i$ and $R_j$, we delete them in index to satisfy the constraint in Definition 6, i.e. two values belongs to different records.

3) *update:* For each value contained in $R_i$ or $R_j$, we update its label in the index and adjust the order of corresponding value pair to satisfy the second constraint in Definition 6.

Please note that merging two records would simplify the index, which could potentially accelerate the process of candidate generation and index maintenance in further entity resolution processing. Such simplification would not affect the correctness of *candidate generation* and *merge* operations in later iterations. We highlight this property as follows.

*Proposition 3:* $\forall i, j \in [1, |\mathcal{R}|]$, let $f(i) =$ find$(i)$, $f(j) =$ find$(j)$, $\mathcal{V}_{f(i)f(j)} \subseteq \mathcal{V}$ always holds.

For any record $R_i$, $f(i)$ is the $rid$ of its corresponding super record. Proposition 3 indicates that for two super records which are merged by arbitrary records, their similar value pairs could be obtained through index, thus ensuing the correctness of candidate generation and merge operation.

*Example 5:* We consider merging $r_1$ and $r_6$ shown in Fig 2. We assume that 1=union(1,6), i.e. $R_1 = r_1 \oplus r_6$.

First, we delete value pairs between $r_1$ and $r_6$. As shown in Fig 6, four value pairs are removed. Next, for each value contained in $r_1$ or $r_6$ (all the $rids$ of such values are in bold), we update its label. Take the value electronics with label $(6, 5, 1)$ for example. After merging into $R_1$, its label updates to $(1, 5, 1)$.

*Proposition 4:* The computational complexity of index maintenance for merging two super records is $O(|\widehat{\mathcal{V}}_{ij}|log|\mathcal{V}|)$.

$|\widehat{\mathcal{V}}_{ij}|$ is number of value pair w.r.t $R_i$ or $R_j$ in index. (In Example 5, it is 9) And the proof is shown in [3].

## IV. VERIFICATION

In this section, we compute record similarity to verify candidate record pairs as discussed in Section III. To achieve this goal, we proposed *instance-based* (Section IV-A) and *schema-based* methods (Section IV-B), respectively.

### A. Instance-based Method

In absence of schema matchings, we treat each record as a set of values ignoring the information of attributes. If two records are similar, they must share some similar values. We compute record similarity by detecting and quantifying the similar degree of value pairs between records, which is captured by *instance-based* method.

For instance-based method, following two problems are to be solved.

1) How to compute record similarity accurately without the information of schema matchings?

2) How to execute computation efficiently due to the sophisticated structure of super record?

Given two records $R_i$, $R_j$ and a value similarity threshold $\xi$, according to Definition 5, first we need to find a field matching set $\mathcal{F}_{ij}$, and then accumulate the similarity values of each field pair covered by $\mathcal{F}_{ij}$ to obtain $Sim(R_i, R_j)$.

Recall that a field matching set (see Definition 4) consists of such field pairs that satisfy the following two conditions: (1) their field similarity is no less than $\xi$; (2) they refer to the same attribute of an entity. We call a field pair as a *similar field pair*, if its similarity is higher than or equal to $\xi$.

Thus, we proceed the process as two steps for record similarity computation in instance-based method, find field matching set and then compute similarity.

**Step 1. Find field matching set**

According to Definition 4, basically, we first need to find all field pairs whose similarity is above $\xi$.

To reach the goal, a naive approach *nest-loop* (shown in Fig 7-(a)) needs to compute the similarity of all value pair $(v_{p,k}^{R_i}, v_{q,l}^{R_j})$ and then calculate field similarity for each field pair. Such a process requires four loops on variable $p, k, q, l$, which is time-consuming.

Recall that a field similarity (see Definition 3) is essentially a value similarity. Instead of nest-loop, we present an efficient algorithm inspired by *similarity join*. We view values in $R_i$ and $R_j$ as two value sets, respectively. Similarity join is performed on those two sets to find all value pairs whose similarity exceeds $\xi$. Based on similar value pair set, we can obtain corresponding *similar field pairs* very fast. In our approach, we do not actually execute similarity join for each record pair, which is quite expensive. Instead, we can obtain the similar field pairs of a record pair leveraging index directly as follows.

Recall that in Section III-B1, $\mathbb{V}_{ij}$ is the result of similarity join on $\mathbb{V}_i$ and $\mathbb{V}_j$, which are the set of values in $R_i$ and $R_j$. $\mathcal{V}_{ij}'$ is refined field set of $\mathcal{V}_{ij}$: for each field pair, keep the value pair with the maximum similarity and delete others, which corresponds to the definition of *field similarity*. (see Definition 3) Thus, $\mathcal{V}_{ij}'$ is exactly the set of all *similar field pairs*. By Algorithm 1 (line 1-8) we can obtain $\mathcal{V}_{ij}'$ through index within $O(|\widehat{\mathcal{V}}_{ij}| + log|\mathcal{V}|)$ steps. (the complexity analysis is presented in [3])

As an example shown in Fig 7, $R_1 = r_1 \oplus r_6$, $R_2 = r_2 \oplus r_4$, we seek to compute $Sim(R_1, R_2)$. Each field is modeled as a point. The dotted line in Fig 7-(a) represents each comparison of field pair by nest-loop. The solid line in Fig 7-(b) expresses the *similar field pair*. From here we observe that *similarity join* reduced comparisons significantly.

After *similar field pairs* are prepared, we seek to select some pairs to generate a field matching set. Next we propose a graph-based method to achieve this goal.

*A Graph-based method*

The pairs in the field matching set must follow one-to-one matching, i.e. they could not share any field. The one-to-one matchings between two field sets may have multiple choices. We observe that a potential true field matching set would maximize $\sum_{(f^{R_i}, f^{R_j}) \in \mathcal{F}_{ij}} simf(f^{R_i}, f^{R_j})$. To achieve the goal, we formulate the problem of deciding field matching set into finding a maximum weighted matching in a bipartite graph as follows.

*Definition 8:* (FIELD MATCHING PROBLEM) Given two records $R_i$, $R_j$ and a value similarity threshold $\xi$, we construct an undirected graph $G(V, E)$, where $V = X \cup Y$ with $X \cap Y = \emptyset$ and $E \subseteq X \times Y$ as follows:

Let a field be a node and a field pair be an edge.

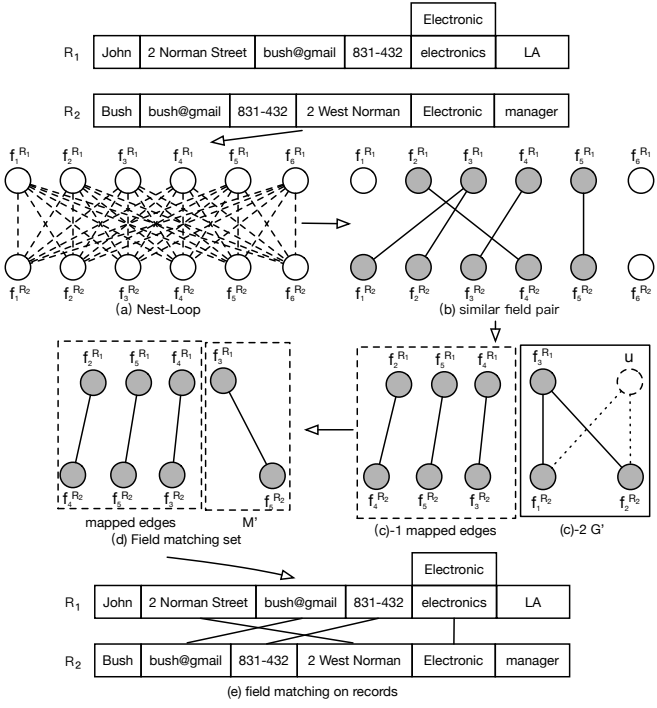

Fig. 7. Instance-based method

1) $E = \{(f_k^{R_i}, f_l^{R_j}) \mid simf(f_k^{R_i}, f_l^{R_j}) \geq \xi\}$;

2) Let $V$ be the point set covered by $E$. $X = \{f_k^{R_i} \mid f_k^{R_i} \in V, 1 \leq k \leq |R_i|\}$ and $Y = \{f_l^{R_j} \mid f_l^{R_j} \in V, 1 \leq l \leq |R_j|\}$;

3) Each edge $e_{kl}$ has a weight $w_{kl}$, which corresponds to $simf(f_k^{R_i}, f_l^{R_j})$.

Then, we aim to find a matching $\mathcal{M}$ of maximum weight, i.e. to maximize $w(\mathcal{M}) = \sum_{e_{ij} \in \mathcal{M}} w_{ij}$.

Before discussing the solution, the following observations lead to chances of simplifying the graph, which can accelerate matching process significantly. For an edge $e$, if the degrees of two vertices of $e$ are both one, then $e$ is deleted from graph. We refer to such an edge $e$ as a *mapped edge*, denoting that two vertices of $e$ are decided to map and thus $e$ can be removed to avoid further computation. Here we denote the degree of a point $u$ by $d(u)$.

*Graph Simplification*

When constructing graph $G$, we count $d(u)$ for each $u \in V$. For each edge $e = (x, y)$, if $d(x) = 1$ and $d(y) = 1$, then we delete $e$ and its endpoints $x$, $y$ from $G$, i.e. $E = E \setminus \{e\}$, $X = X \setminus \{x\}$ (if $x \in X$), $Y = Y \setminus \{y\}$ (if $y \in Y$).

Let $G' = (X' \cup Y', E')$ be the simplified graph, and $\mathcal{E}$ be the set of deleted edges, i.e. *mapped edges*. As shown in Fig 7, Fig 7-(c)-1 represents the mapped edges. The solid points and lines in Fig 7-(c)-2 denotes $G'$.

So far, the field matching problem has been converted to finding a *Maximum Weight Matching* on $G'$, which can be solve by a well-known Kuhn CMunkres(KM) algorithm [11] (details of KM are shown in [3]. Note that KM algorithm requires a complete bipartite graph, that is, $|X'| = |Y'|$. We can add dummy points and set the weight of their corresponding edges to be zero to satisfy above condition. As shown in Fig 7-c, we add $u$ to let $|X'| = |Y'|$ and the weight of edges connecting with $x$ is 0. The output of KM on $G'$ is a matching $\mathcal{M}'$ with maximum weight.

Finally, we have obtained $\mathcal{M}'$ and a set of mapped edge $\mathcal{E}$, which is shown in Fig 7-d. Next we show that $\mathcal{M}' \cup \mathcal{E}$ is the match of maximum weight in graph $G$.

*Theorem 1:* The match of maximum weight in $G$, $\mathcal{M} = \mathcal{M}' \cup \mathcal{E}$.

Theorem 1 indicates that the deleted edges are a part of optimal solution, and the simplification of graph will not affect the optimality of solution. We show the proof in [3]. According to Definition 8, the field matching set of $R_i$ and $R_j$ corresponds to $\mathcal{M}$.

**Step 2. Similarity Computation** Finally, we accumulate similarity values of field pairs in $\mathcal{F}_{ij}$ and compute $Sim(R_i, R_j)$ according to Definition 5.

The time complexity of *instance-based* method is $O(log|\mathcal{V}|) + O(m^3)$ ($m = \max(|X'|, |Y'|)$), where $O(log|\mathcal{V}|)$ is computational steps of finding $\mathcal{F}ij$ through index and $O(m^3)$ is contributed by KM algorithm. Note that, although the complexity is cubic w.r.t. $m$, our experimental results in Section VI show that in many real-world datasets $m^3$ is far less than $|\mathcal{V}|$ and *instance-based* method is very fast in practice.

The *instance-based* method has two results: record similarity and predicated schema matchings produced by KM algorithm (shown in Fig 7-d). Each time if two records are judged to be similar, then they introduce some new schema matching information. The next part *schema-based* method tends to leverage such information to decide the true schema matchings, which can accelerate similarity computation and improve accuracy.

*B. Schema-based Method*

*Schema-based* method aims to collect the schema matching information produced by *instance-based* method to decide potential true schema matchings.

We observe that schema matching predictions provided by *instance-based* method for various record pairs may contradict each other. Consider motivating example in Fig 1. We set $\xi = 0.3$, John in $r_1$ is similar to john@bush in $r_5$, indicating that Customer I.name≈Customer III.e-mail. However, John in $r_1$ is identical with John in $r_6$, indicating that Customer I.name≈Customer III.name. For Customer I.name, we aim to determine the true matching among different providers. Under the assumption[12] that there are no redundant attributes in one schema, for an attribute, among its predicted matching attributes under one schema, only one is true.

Under schema $s_i$, for an attribute $a_k^i$, there are $n$ predictions so far. We formulate $n$ predications as $n$ independent trials. Let $x_i$ be the outcome of the $i^{th}$ trial and $x^*$ be the true matching. Then we construct independent Bernoulli random variables $X_1, X_2, ..., X_n$ as follows: if $x_i \neq x^*$, let $X_i = 0$; if $x_i = x^*$, let $X_i = 1$.

Intuitively, among $n$ trials, if a matching occurs more often, then it could more likely be the true one. Thus, we adopt a majority vote method to estimate the true matching. Specifically, among $n$ trials, we take the outcome with the highest frequency as the true one. However, such a vote method may lead to errors, especially on the case when $n$ is small. Accordingly, we derive the upper bound of error probability, denoted by $UP_{error}$.

Let $\widehat{x}$ be the estimated value by majority vote. We denote by $p = Pr(x = x^*)$, where $p$ is a priori value obtained by training dataset. Then we give the upper bound as follows.

*Theorem 2:* $UP_{error} = e^{-\frac{n}{2p}(p-\frac{1}{2})^2}$

We show the proof of Theorem 2 in [3].

Given a error probability threshold $\rho$, if $UP_{error} < \rho$, we decide $\widehat{x}$ to be the true matching with the probability of $1 - UP_{error}$ and embed it into *instance-based* method in the following comparisons. That is, once a matching $(a_k^i \approx a_l^j)$ is determined to be true, in the later comparisons we can directly include corresponding field pair $(f_k^{R_i}, f_l^{R_j})$ into the field matching set.

As an example, suppose $p = 0.8, n = 10, \rho = 0.6$. We have $UP_{error} = 0.57$ and we decide $\widehat{x}$ as the true matching with the probability 0.43.

## V. OVERALL SOLUTION

We introduce our ***Heterogeneous Entity Resolution Algorithm*** in Algorithm 2.

**Algorithm 2** HERA
**Input:** Record set $\mathcal{R}$,
  record similarity $\delta$, string similarity $\xi$;
**Output:** $\mathcal{R}' : \{(r_i, eid) \mid r_i \in \mathcal{R}\}$ Record set with entity labels;
1: *build index*;
2: **while** !*stop_condition* **do**
3:   *Candidate generation*;
4:   **for** each $(R_i, R_j) \in \mathbb{R}'$ **do**
5:     $merge(R_i, R_j)$;
6:   **for** each $(R_i, R_j) \in \mathbb{R}$ **do**
7:     Compute $Sim(R_i, R_j)$;
8:     **if** $Sim(R_i, R_j) \geq \delta$ **then**
9:       $schema\_based(R_i, R_j)$;
10:       $merge(R_i, R_j)$;
11: **for** $r_i \in \mathcal{R}$ **do**
12:   $\mathcal{R}' \leftarrow \mathcal{R}' \cup (r_i, \text{find}(i))$;
13: **return** $\mathcal{R}'$;

Initially, we build index as Definition 6 (Line 1). The stop_condition is that $\forall R_i, R_j \in \mathcal{R}, Sim(R_i, R_j) \geq \delta$. When no two records can be further merged, our algorithm stops (Line 2). Next, we perform Candidate generation to decide candidate record pairs, which is described in Section III-B1 (Line 3). During candidate generation, besides pruning dissimilar record pairs, we also compute the similarities of a record set directly, to generate $\mathbb{R}'$. For each record pair in $\mathbb{R}'$, we merge them directly (Line 4-5). Next for each record pair in candidates set $\mathbb{R}$, we compute record similarity (using instance-based method in Section IV-A). If $R_i$ and $R_j$ are similar, we execute schema-based method to use the predictions of schema matchings to decide true ones. (see Section IV-B). Then we merge two record using techniques in Section III-B2 (Line 6-10). Finally, for each record $r_i$, we use union-find to get the rid of its corresponding super record, which is used as its entity label (Line 11-12).

Continue with motivating example. We set value similarity threshold $\xi = 0.5$ and record similarity threshold $\delta = 0.5$. $\mathcal{R} = \{r_1, r_2, ..., r_6\}$ and Fig 8 shows the overall process of HERA. In Iteration 1, we merge three similar record pairs into three super records $R_1$, $R_2$ and $R_3$, respectively. It is worth to note that the index is updated arisen from merge operation. In iteration 2, we consider three super records $\{R_1, R_2, R_3\}$ instead of $\{r_1, ..., r_6\}$. Since only the similarity of $(R_1, R_2)$ is above 0.5, we merge them into a new super record $R_1$. (assume that 1=unoin(1,2)) Finally, the stop condition is satisfied. $r_1, r_2, r_4, r_6$ are merged into super record $R_1$, thus they refer to the same entity. Similarly, $r_3$ and $r_5$ refer to the same entity.

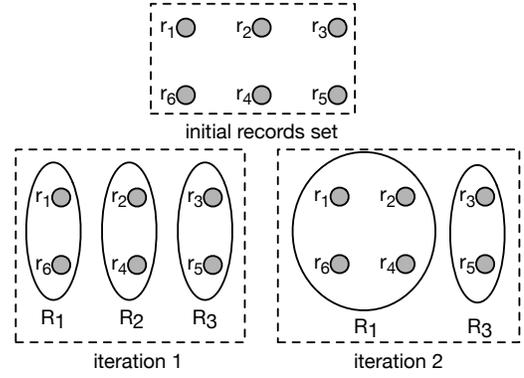

Fig. 8. Overall solution

| | $D_{m1}$ | $D_{m2}$ | $D_{m3}$ | $D_{m4}$ |
|---|---|---|---|---|
| $n$ | 1000 | 2000 | 3000 | 4000 |
| # of entity | 121 | 277 | 361 | 533 |
| # of distinct attribute | 16 | 22 | 23 | 21 |

TABLE I.  TECHNICAL CHARACTERISTICS OF $D_{m1}$-$D_{m4}$

*Proposition 5:* Time complexity of *HERA* is $O(k(|\mathcal{V}| + |\mathbb{R}|(\overline{m}^3 + |\overline{\mathcal{V}}|log|\mathcal{V}|)))$.

$k$ is the number of iterations, $|\mathbb{R}|$ is the number of candidate record pairs, $\overline{m}$ is the average number of points in the simplified bipartite graph, and $|\overline{\mathcal{V}}|$ is the average number of value pairs covered by any compared record pairs in index. The proof is shown in [3].

## VI. EXPERIMENT

In this section, we report the experimental evaluation on comparing our method with the state-of-the-art and testing performance under different parameter settings.

### A. Experimental Setup

**Data Sets.** We employed a real data set $D_{movies}$ in [15], [1], which is a collection of movies shared among IMDB and DBPedia. [1] To derive the ground truth, we relied on the "*imdbid*" attribute in the profiles of the DBPedia movies. Since $D_{movies}$ is pretty large and the focus of our work is not scalability, we extracted four relatively small data sets from $D_{movies}$, $D_{m1}$, $D_{m2}$, $D_{m3}$ and $D_{m4}$, respectively.

Table I lists the technical characteristics $D_{m1}$-$D_{m4}$. The number of entities could be counted by ground truth. Whether two attributes are distinct is distinguished manually, thus the number of distinct attributes can be obtained.

Since most state-of-art focus on records under a predefined schema (as mentioned in related work), in order to make the experiment comparable, we conduct it as follows: we perform *schema matching* and *data exchange* on the above four datasets to generate another 8 datasets. Next we take $D_{m1}$ as an example. We generated two datasets $D_{m1-S}$ and $D_{m1-L}$ from $D_{m1}$ as follows.

*1) Schema matching*: Let $A_1$ be the number of distinct attributes contained in $D_{m1}$. Whether two attributes are matched could be distinguished manually. Since in most scenarios a target schema is user-defined and created for individual goals, we randomly selected part of distinct attributes from source schemas contained in $D_{m1}$ to generate the target schema. To vary the information content between source schemas and target schema, we randomly chose $A_1/3$ and $2*A_1/3$ distinct attributes in $DM1$ to form the target schemas of $D_{m1-S}$ and $D_{m1-L}$, respectively. Then we decide the schema matchings manually.

---
[1] http://wiki.dbpedia.org/Datasets

|   | $D_{m1}$ | $D_{m2}$ | $D_{m3}$ | $D_{m4}$ |
|---|---|---|---|---|
| $|\mathcal{S}|$ | 13294 | 39270 | 52463 | 79462 |
| $\overline{m}$ | 8.3 | 11.2 | 7.9 | 8.6 |
| $k$ | 19 | 24 | 27 | 26 |

TABLE II. PARAMETERS FOR DIFFERENT DATASETS

2) *Data exchange*: Following [16], we transformed the schema matchings to corresponding tgds (*tuple generating dependency*) and used the data exchange modeling tool [17] to convert instances in $D_{m1}$ to $D_{m1-S}$ and $D_{m1-L}$.

The same operations are executed on $D_{m2}$, $D_{m3}$, $D_{m4}$ to generate $D_{m2-S}$, $D_{m2-L}$, $D_{m3-S}$, $D_{m3-L}$, $D_{m4-S}$, $D_{m4-L}$, respectively. In the remainder of our paper, we call such 8 generated datasets as *homogeneous* datasets.

**Implementation.** We build index for $D_{m1}$-$D_{m4}$ and leave the details in [3].

Our experimental evaluations mainly consist of two aspects, *efficieny* and *effectivness*.

Regarding efficiency, we first examined some key parameters for $D_{m1}$-$D_{m4}$: $\mathcal{S}$, the size of index; $\overline{m}$, the average number of points in simplified bipartite graph described in Section IV-A; and $k$, the number of iterations. Then we reported run time and the number of comparisons of HERA conditioned on various thresholds on $D_{m1}$-$D_{m4}$.

Regarding effectiveness, we tested the quality of results returned by HERA on $D_{m1}$-$D_{m4}$. Next, we compared HERA with the other three state-of-art algorithms, R-Swoosh [4], CC [6] (Correlation clustering) and CR[5] (Collective ER) on 8 homogeneous datasets.

In this section, we report the quality of results on $D_{m1-S}$ to $D_{m4-S}$. The results on the other four datasets are shown in [3].

We implemented the algorithms in C++ on a Windows machine with Intel Core i5 processor.

**Measure:** We compared the results of our method with ground truth and measured its quality by *precision*, *recall* and *F1-measure*. *Presicion* is defined as the proportion of correctly identified record pairs to the record pairs generated by HERA. *Recall* is the proportion of correctly identified record pairs to the correct record pairs based on the ground-truth entities, and F1-measure is the harmonic average of precision and recall: 2/(1/*precision* + 1/*recall*).

### B. Efficiency

We conducted three experiments to show the performance of HERA. Table II lists $|\mathcal{S}|$, $\overline{m}$ and $k$ on four datasets. The average edges of simplified bipartite graph $m$ is a fairly small number, indicating that the simplification approach reduces graph size significantly. In contrast, the number of value pair in index is a considerably large one, up to around 80 thousand in $D_{m4}$.

Fig 10(a) shows the number of comparisons with respect to different thresholds. On these four datasets, we have the similar observations: as we increase $\delta$, the number of comparisons in HERA declines. The decrease can be attributed to the fact that a higher threshold reduces the size of candidate record pairs, leading to less comparisons.

Fig 12(a) depicts the execution time on different datasets. As performing HERA on datasets with larger size, it requires more time. When $\delta$ increases, the time cost on different datasets varies more slightly. Typically, as $\delta$ is 0.8, HERA finished in about 100 ms for all datasets.

### C. Effectiveness

In this part, we first conducted experiments to show the quality of results on four heterogeneous datasets. Next we compared with some state-of-art approaches, R-Soosh, CR and CC on datasets $D_{m1-S}$, $D_{m2-S}$, $D_{m3-S}$ and $D_{m4-S}$.

To examine how threshold affects precision, we report the comparison results in Fig 9(a). Generally, we observe a slight decline for precision as the size of datasets expanding. In contrast, the decline turns pronounced when $\delta$ drops. It decreases by 9% on average from $D_{m1-S}$ to $D_{m4-S}$. In Fig 9(b), we plots recall w.r.t. different thresholds. The curves climb dramatically as $\delta$ increases. On $D_{m1-S}$, recall is 0.98 when $\delta = 1$ and it turns 0.81 when $\delta = 0.2$. Regarding the size of datasets, similarly, the larger size, the lower recall. On $D_{m4-S}$, the worst recall is 0.72 ($\delta = 0.2$) while the worst recall on $D_{m1-S}$ turns 0.81. F1-measure is presented in Fig 9(c), and we have the following observations. First, F-measure increases till the best value and then drops slightly. The best value for $D_{m1-S}$ is 0.8, and for others is roughly 0.6. Second, as we queried more records, F-measure drops slightly. The average F-measure on $D_{m1-S}$ is 91.6, on $D_{m4-S}$ is 86.4, decreasing by 4.4%.

Next we compared the accuracy of HERA with R-Soosh, CR and CC in terms of precision, recall and F-measure. Fig 11(a) shows the precision. Compared with three competitors, HERA has the best performance. Specifically, the average precision exceeds 0.9, which improves R-SWoosh by 6%, CR by 12% and CC by 13%. Among them, CR and CC have similar precision on four datasets. Considering recall shown in Fig 11(b), HERA also outperforms others on all datasets. The average recall of HERA is 92.75, beats R-Swoosh by 6%, CR by 10% and CC by 16%. CR and CC are slightly sensitive to the size of datasets and CC obtains a low recall below 0.8 on four datasets. On datasets with relatively small size, R-Swoosh has competitive performance. Its recall rate reaches nearly 0.9. Finally, regarding F-measure, HERA obtains a significant improvement in comparison to the other methods. On average, it outperforms R-Swoosh by 6%, CR by 11%, CC by 15%. Furthermore, HERA is insensitive conditioned on datasets with different size (range roughly from 1000 to 4000) while the F-measure of other methods shows a pronounced decline. Thus, we can draw that HERA could collect more information to improve ER on datasets with homogeneous schema.

## VII. RELATED WORK

A variety of methods for solving the ER problem have been proposed in literature (surveyed in [18], [19]. Most prior work in this area has focused on ER on records under a predefined schema.[2] Only a few work study ER under heterogeneous settings.

[9], [10] used learning-based methods to solve ER. Among them, [9] applied classification techniques drawn from statistical pattern recognition, machine learning, etc. to decide

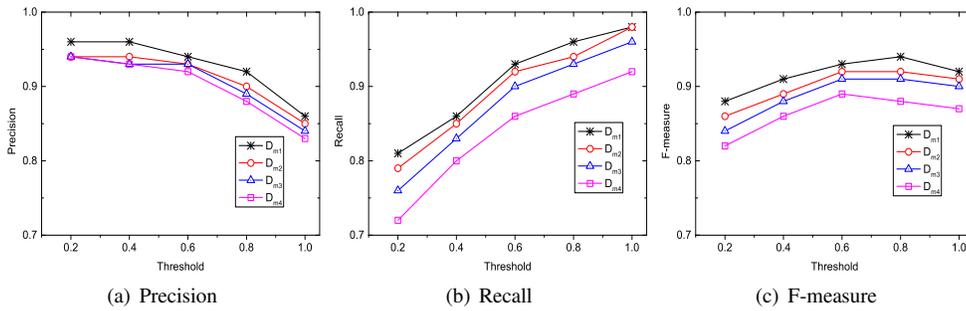

Fig. 9. Effectiveness test on $D_{m1}$ to $D_{m4}$

(a) Precision  (b) Recall  (c) F-measure

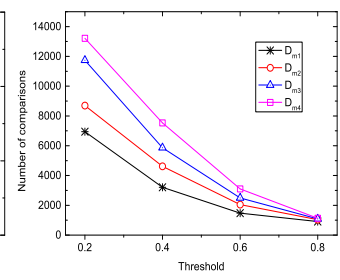

Fig. 10. # of comparisons

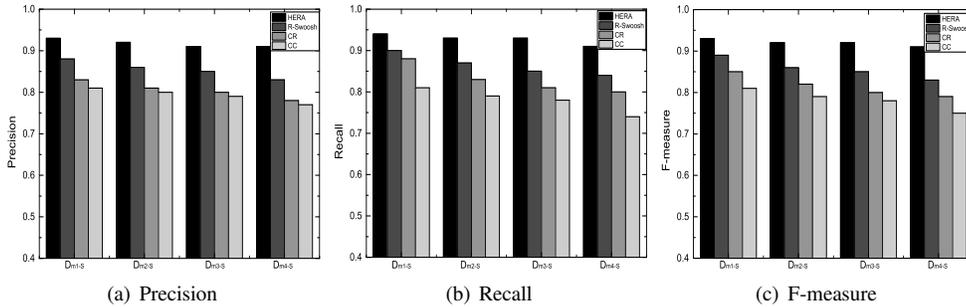

Fig. 11. Effectiveness test on $D_{m1-S}$ to $D_{m4-S}$

(a) Precision  (b) Recall  (c) F-measure

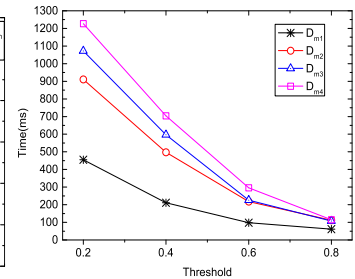

Fig. 12. Runtime

whether two records represent the same real-world entity. [10] presented a algorithm *HFM* that combines the machine learning and expert systems approaches to determine true field matchings. However, both above methods fail to recognize those record pairs satisfying *description difference*.

Some other work [1] focused on the blocking techniques of ER under heterogeneous environment. They introduced the attribute-agnostic blocking methodology and made no use of schema information in the blocking step of ER, which could be applicable in heterogeneous settings with loose schema binding. Their work did not comprise the exact solution of record similarity computation and such a blocking technique still failed to satisfy the *description difference*.

Also some human-based approaches [7], [8] could be used to solve heterogeneous ER. They try to assign some HITs to human workers. However human-based method of ER in heterogeneous settings falls into low effectiveness since records satisfying *description difference* are difficult recognized by human. (see $r_1$ and $r_2$ in Fig 1)

As for the work studying ER on records under the same schema [4], [5], [6], an inherent shortcoming is the loss of information, depending on the predefined schema. To make up this shortcoming, we propose HERA for heterogeneous datasets and perform data exchange to convert records to those under a given schema.

## VIII. CONCLUSION

In this paper, to solve ER under heterogeneous environment, we propose HERA. Basically, we describe a *merge-and-compare* mechanism to solve *description difference*. Regarding efficiency, we design an index to generate candidates and speed up record similarity computation. Without the information of schema matchings, we present instance-based and schema-based method to verify the candidates. We show that two methods can benefit each other to accelerate similarity computation and improve the accuracy. Experimental results show that our methods can significantly improve the state-of-art even on records under a predefined schema.